\documentclass[pdflatex,sn-mathphys]{sn-jnl}% Math and Physical Sciences Reference Style
\jyear{2021}%

\theoremstyle{thmstyleone}%
\theoremstyle{thmstyletwo}%

\theoremstyle{thmstylethree}%

\begin{document}

\title[Nanomagnetism in porous amorphous palladium]{Nanomagnetism in porous amorphous palladium, a sequel. A possible light-weight magnet.}


\author[1]{\fnm{Isaías} \sur{Rodríguez}}\email{isurwars@gmail.com}

\author[2]{\fnm{Renela M.} \sur{Valladares}}\email{renelavalladares@gmail.com}

\author[1]{\fnm{David} \sur{Hinojosa-Romero}}\email{david18\_hr@ciencias.unam.mx}

\author[2]{\fnm{Alexander} \sur{Valladares}}\email{valladar@ciencias.unam.mx}

\author*[1]{\fnm{Ariel A.} \sur{Valladares}}\email{valladar@unam.mx}

\affil[1]{\orgdiv{Instituto de Investigaciones en Materiales}, \orgname{Universidad Nacional Autónoma de México}, \orgaddress{\street{Apartado Postal 70-360}, \city{Ciudad Universitaria}, \postcode{04510}, \state{CDMX}, \country{México}}}

\affil[2]{\orgdiv{Facultad de Ciencias}, \orgname{Universidad Nacional Autónoma de México}, \orgaddress{\street{Apartado Postal 70-542}, \city{Ciudad Universitaria}, \postcode{04510}, \state{CDMX}, \country{México}}}


\abstract{Magnetism is a very relevant subject that permeates our everyday lives. However, magnetism keeps taking us from surprise to surprise which seems to indicate that it is a phenomenon not well understood. For example, we found that bulk amorphous palladium becomes magnetic; so, naturally one should ask, could defective palladium develop magnetism? In particular, would amorphous porous palladium become magnetic? Here we show that the answer to that question is affirmative, this defective topology of Pd is magnetic, with a magnetism that depends on the amount of sample porosity and on the topology of their structures. Clearly, if magnetism exists in porous amorphous palladium, this indicates the possibility of developing light-weight magnets, useful when a maximized magnetism/weight ratio is demanded, well suited for space and aeronautical applications. }




\maketitle

\section{Introduction}\label{Introduction}
Porous science is in the making. Porous materials have existed before humanity, but a systematic and fundamental approach to its understanding is recent. This explains the limited number of papers that deal with the basics of the subject. We pretend to contribute to this understanding by performing computational simulations of the topology, the electronic structure and the magnetic properties of some porous and amorphous metallic systems, pure and alloyed. The present study follows on the footsteps of our prediction that amorphous palladium, $a$-Pd, displays magnetism in the bulk \citep{rodriguez_emergence_2019}. Our reasoning is that since a defective structure of palladium, like the amorphous phase, displays magnetism then an even more defective one, like the porous-amorphous topology, $pa$-Pd, should display magnetism as well. If so, this could become an important asset in the technological world since it would be possible to have light weight magnets. \\

Palladium behaves in a non-expected manner when expanded or contracted \citep{moruzzi_magnetism_1989}, or confined to a restricted dimensionality \citep{sampedro_ferromagnetism_2003} and certainly when amorphous in the bulk \citep{rodriguez_emergence_2019}, as opposed to the behavior of the lone atom. It has also been found that Pd develops magnetism when alloyed with H \citep{rodriguez_superconductivity_2020, mott_theory_1958}, or silicon, both liquid \citep{muller_magnetic_1978} or solid \citep{rodriguez_ab_2020}. Moruzzi and Marcus \citep{moruzzi_magnetism_1989} studied the electronic properties of crystalline Pd and Rh from below the equilibrium volume to an expanded state to simulate free atoms, and they discovered that palladium, like most condensed transition metals, must also exhibit magnetic behavior at sufficiently large volumes, although the free lone atom of Pd is not a magnetic entity but acquires the $4d^{10}5s^{0}$ structure. The expectation that bulk crystalline Pd could be magnetic even at equilibrium volumes has led to a sustained interest in its electronic structure and, consequently, its magnetic properties.\\

Porous palladium, $p$-Pd, was first reported by Kabius, Kaiser and Kaesche (KKK) \citep{kabius_micromorphological_1986} in 1986. KKK produced $p$-Pd by selective chemical dissolution of copper in a CuPd alloy. However, their accomplishment remained archived until 2009 when Balk and coworkers resumed its study and produced nanoporous Pd thin films by a selective dissolution, dealloying, of PdNi alloy films in dilute sulfuric acid \citep{li_preparation_2009}. Their objective was to investigate the hydrogen absorption/desorption behavior in this hydride, but, in the process they were able to generate structures that were sponge-like with ligaments/pores $\sim$10 nm. depending on the experimental conditions. Dealloying nickel was an incomplete process since the final result was a 46 at. \% nickel remnant, although the XRD scans did not detect the PdNi alloy peak \citep{li_achieving_2010}. They studied the topology of films that were 90 nm thick, with an initial concentration of 18 at. \% Pd and 82 at. \% Ni, and they found finer pores and ligaments in the final sample. These pores and the ligaments are larger than the ones reported in this work as shall be shown later. Selective dissolution, or dealloying, is a process used by KKK to create porous palladium by dissolving copper in a PdCu alloy, and it generates large stresses in the material and the structure created is in a metastable state that would evolve into a different, stable, structure when equilibrium is reached. Nevertheless, lacking an experimental method to expand (negative pressures) solid matter, this dealloying approach may become an interesting route to produce porous materials.\\

Other works on the experimental generation of porous palladium are found in references \citep{hakamada_preparation_2009,yu_nanoporous_2008,guo_synthesis_2008}, all since 2008. However, to the best of our knowledge, no reference to magnetic properties of porous-amorphous palladium has been made; therefore, we decided to extend the investigation of the magnetism found in $a$-Pd \citep{rodriguez_emergence_2019} to enquire into a possible magnetism of a defective structure like $pa$-Pd. In this work we report findings that indicate the presence of nanomagnetism in samples with different porosities (generated using the \textit{expanded- lattice} approach cultivated in our group \citep{santiago-cortes_computational_2012}), and with different structures; some resulting from a molecular dynamics (MD) procedure applied to the \textit{expanded-lattice} structures and some others in local equilibrium, consequence of a geometry optimization (GO) procedure applied to the outcome of the MD process. In both cases magnetism appears and the intensity depends on the structure for a given porosity. Due to the size of the supercells we used, we speak of the magnetism found, as nanomagnetism, or perhaps more appropriately, sub-nanomagnetism, because the dimensions of the “domains” found are in the sub-nano range. The relative orientations determine the final magnetic moment of our supercells.

\section{Methodological considerations}\label{Method}

When dealing with a material as interesting as palladium the implications of a given result should be fully appreciated and extended to study related situations that may occur. Since the magnetism discovered for amorphous bulk Pd may be due to defects present in the samples \citep{rodriguez_emergence_2019} the natural thing to do was to study other defective structures and ponder their possible magnetism. Amorphous alloys could be considered as defective matter and that is why we undertook the study of the solid amorphous palladium “ides”, (Pd$_{1-x}$(H,D,T)$_x$), and found that they developed magnetism as a function of the “ides” concentration \citep{rodriguez_superconductivity_2020}. Also, the study of amorphous solid PdSi alloys showed magnetism as well \citep{rodriguez_ab_2020}. However, the natural extension of our work on amorphous palladium is to generate porous amorphous samples using two approaches that we have developed in our workgroup: the \textit{undermelt-quench} approach \citep{valladares_new_2008} used to generate amorphous materials and the \textit{expanded-lattice} approach \citep{santiago-cortes_computational_2012} instrumented to generate porous samples. The \textit{undermelt-quench} approach has been applied to Pd and to many other materials to generate their amorphous structures (see Refs. \cite{rodriguez_emergence_2019} and \cite{valladares_new_2008}). The \textit{expanded-lattice} approach has also been applied to lattice sizes manageable with first-principles methods, since these methods are very computer-intensive and demand vast amounts of computational resources. This limitation restricts us to using supercells with a number of atoms of the order of a few hundred, and therefore the apparent existence of magnetic regions, found in this work, smaller than the well-known magnetic domains, (normally of the order of nanometers), should be studied more extensively and intensively. We pursued our goal, nevertheless, and generated supercells that contain 216 atoms with lattice parameters given in Table \ref{tab1}.

\bigskip
\begin{table}[ht]
\begin{center}
\begin{minipage}{200pt}
\caption{Porosities, lattice parameters, and densities of the supercells studied in this work.}\label{tab1}%
\begin{tabular}{@{}lccc@{}}
\toprule
Name & Porosity  & Lattice Parameter & Density\\
 & (\%) & [a=b=c] (\AA) & (g cm$^{-3}$)\\
\midrule
$a$-Pd              & 0     & 14.7058  & 12.00  \\
$pa$-Pd$_{25\%}$    & 25    & 16.1858  & 9.00  \\
$pa$-Pd$_{50\%}$    & 50    & 18.5281  & 6.00  \\
$pa$-Pd$_{62.5\%}$  & 62.5  & 20.3929  & 4.50  \\
$pa$-Pd$_{75\%}$    & 75    & 23.3440  & 3.00  \\
\botrule
\end{tabular}
\end{minipage}
\end{center}
\end{table}
\bigskip

We started by expanding the lattices to generate the porosities mentioned in Table \ref{tab1} and of course the 0\% porosity sample reported in Ref \cite{rodriguez_emergence_2019}. The computer simulations were conducted in CASTEP \citep{clark_first_2005} an \textit{ab-initio} software included in the Materials Studio suite \citep{dassault_systemes_biovia_materials_2016}. 
The parameters used in the simulations are as follows. For both the MD and GO, the PBE exchange-correlation functional was used \citep{perdew_generalized_1996}, with a cut-off energy of 300 eV for the plane-wave basis. The MD process consists of 225 steps with a time-step of 5 fs with an initial increasing linear run of 100 steps starting at a temperature of 300 K and reaching a maximum temperature of 1500 K with a slope of 12 K per step. Then a quenching phase of 125 steps was executed also with a slope of 12 K per step going from 1500 K to a minimum of 12 K, according to the \textit{undermelt-quench} procedure \citep{valladares_new_2008}. Care was exercised to keep the atomic spins free to orient according to the potentials (forces) in the quantum mechanical \textit{ab-initio} computing process. The thermostat used in both thermal processes was Nóse-Hoover within an NVT ensemble. After the MD runs were accomplished we proceeded to optimize the resulting structures to find a (local) minimum of energy by performing a GO. For the energy minimization, the density mixing was used with Pulay Scheme, with a charge mixing of 0.5, and a spin mixing of 2.0. Thermal smearing in the occupancy of 0.1 eV was utilized, and an energy convergence threshold of 5x10$^{-7}$ eV for the SCF cycles. For this process, the BFGS minimizer was selected with a geometry energy tolerance of 2x10$^{-5}$ eV, a force tolerance of 5x10$^{-2}$ eV \AA$^{-1}$, a stress tolerance of 5x10$^{-2}$ eV \AA$^{-3}$, and a max displacement of 2x10$^{-3}$ \AA. \\

We analyzed the topology of the resulting structures using Correlation, a software developed by our workgroup \citep{rodriguez_correlation_2021}, and obtained their electronic and magnetic properties after both the MD and the GO runs.\\

\section{Results}\label{Results}

\begin{figure}[ht]%
\centering
\includegraphics[width=\textwidth]{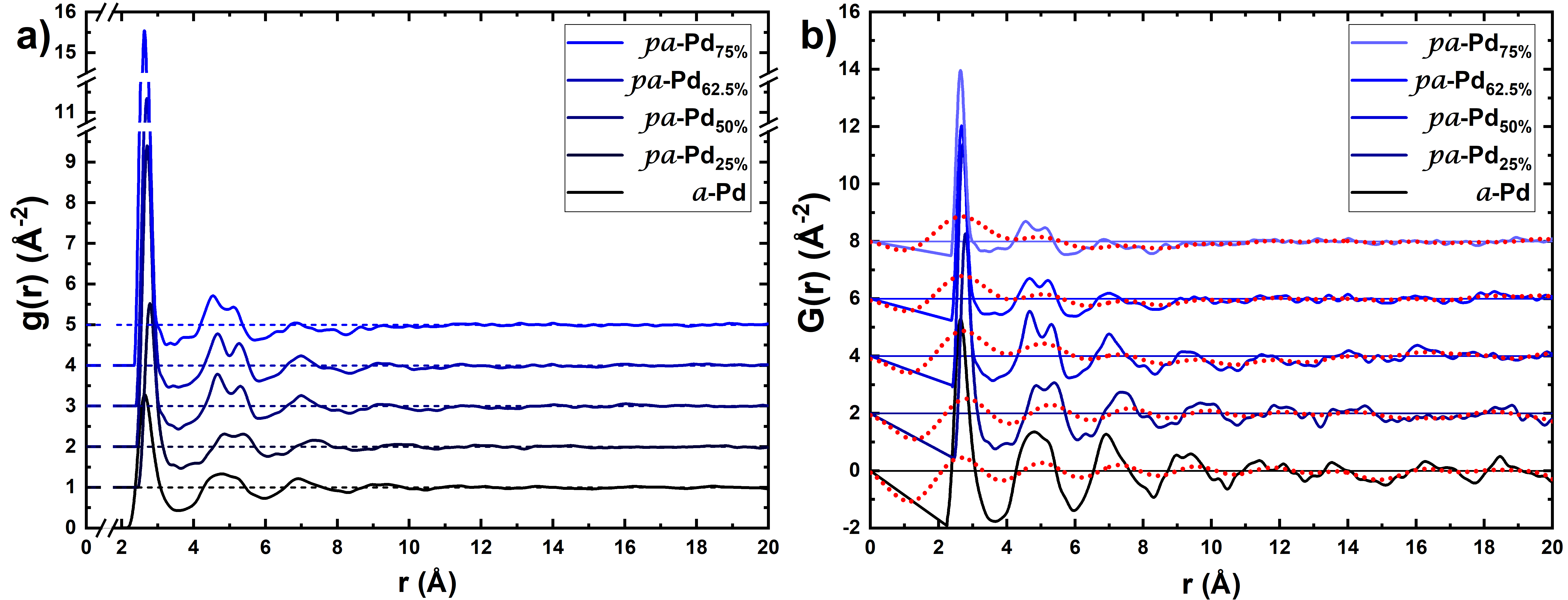}
\caption{a) Pair Distribution Functions and b) reduced Pair Distribution Functions, after the MD processes. Average of the cells of every porosity shown. Color online.}\label{fig1}
\end{figure}

\begin{figure}[ht]%
\centering
\includegraphics[width=\textwidth]{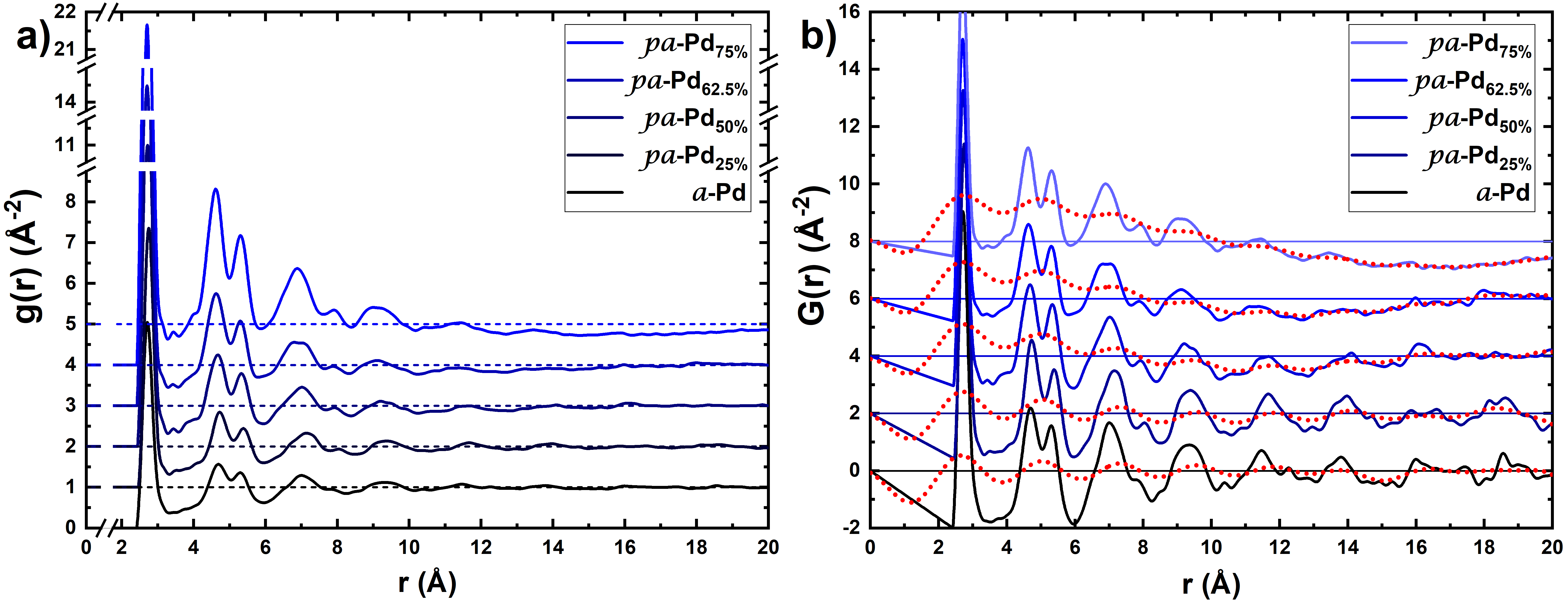}
\caption{a) Pair Distribution Functions and b) reduced Pair Distribution Functions, after the GO processes. Average of the cells of every porosity shown. Color online.}\label{fig2}
\end{figure}

\begin{figure}[ht]%
\centering
\includegraphics[width=\textwidth]{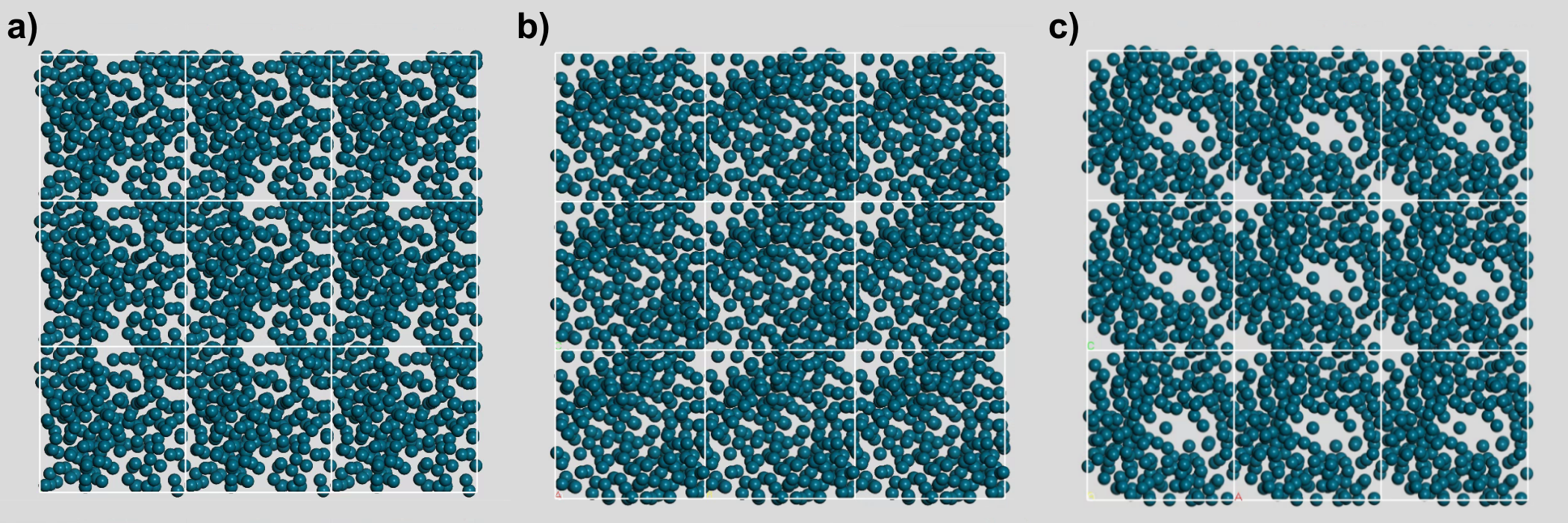}
\caption{Typical topology of a porous structure after the MD run. The porosity displayed corresponds to 50\% and the view is (a) perpendicular to the xy plane, (b) perpendicular to the yz plane, and (c) perpendicular to the zx plane. Color online.}\label{fig3}
\end{figure}

\begin{figure}[ht]%
\centering
\includegraphics[width=\textwidth]{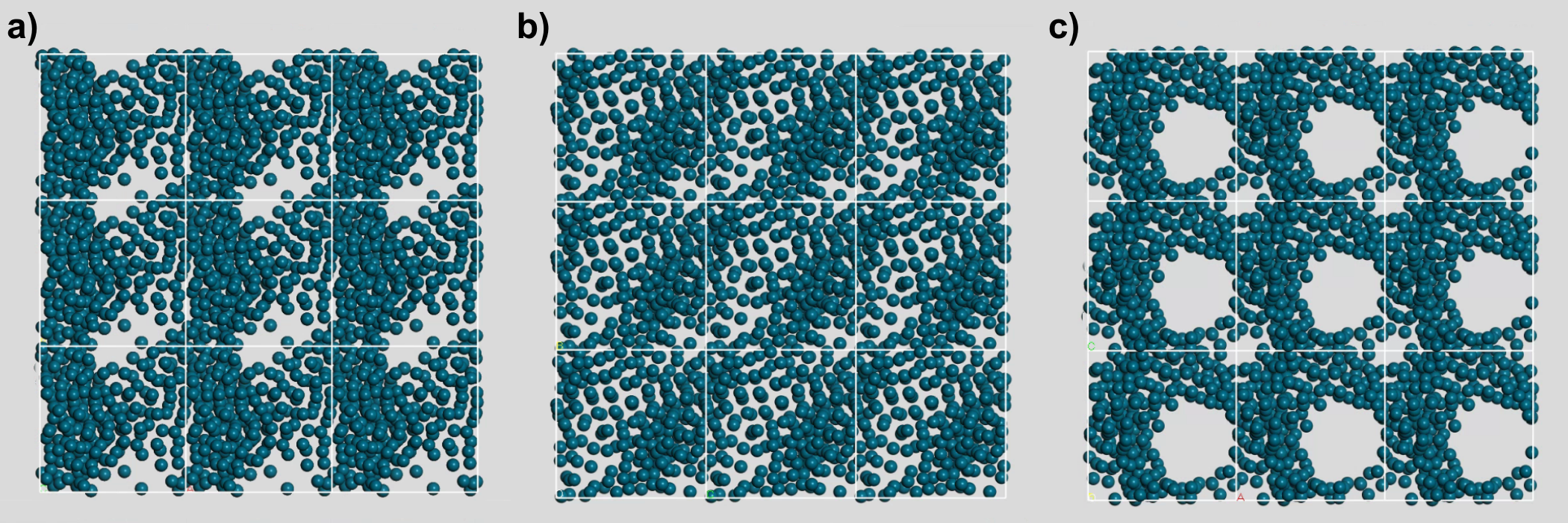}
\caption{Typical topology of a porous structure after the GO run. The porosity displayed corresponds to 50\% and the view is (a) perpendicular to the xy plane, (b) perpendicular to the yz plane, and (c) perpendicular to the zx plane. The difference with Figure \ref{fig3} is to be noted. Color online.}\label{fig4}
\end{figure}

The generated atomic topologies were characterized by obtaining the Pair Distribution Functions (PDFs), $g(r)$, and the reduced Pair Distribution Functions (rPDFs), $G(r)$, at the end of each MD and GO run \citep{musgraves_springer_2019, waseda_structure_1980}; PDFs and rPDFs are shown in Figures \ref{fig1} and \ref{fig2}. In a previous work \citep{santiago-cortes_computational_2012}, we tried to relate the extent of the region where the corresponding PDF has a value below 1, beyond the middle-range order, to the size of the pore and the conclusions were not completely satisfactory. However, this depletion appears again in this work and by \textit{reductio ad absurdum} there must be some relation between this region and the existence of pores, phenomenon to be elucidated. To emphasize this feature, we decided to plot the rPDFs in the corresponding Figures \ref{fig1} and \ref{fig2} by increasing the number of points used in the Fast Fourier Transform (FFT) to smooth the functions and, in fact, the over-smoothed red-dotted curves in the rPDFs indicate the presence of the pores. The regions where the PDFs are less than 1, and where the rPDFs are less than 0, beyond the middle-range order, signal the existence of pores, and a comparison between these values and those obtained by a geometrical estimation, using Connolly surfaces, is given in Table \ref{tab2}. At the end of the MD processes the topology of the structures is more dendritic-like, whereas when the GO is performed large through pores (voids) appear, (See Figures \ref{fig3} and \ref{fig4} where the 50\% porosity structures are shown). 

\begin{table}[ht]
\begin{center}
\begin{minipage}{\textwidth}
\caption{Comparison of the size of the pores obtained by a geometrical estimation and by evaluating the depletion of the $G(r)$.}\label{tab2}
\begin{tabular*}{\textwidth}{@{\extracolsep{\fill}}lcccc@{\extracolsep{\fill}}}
\toprule%
Name & \multicolumn{2}{@{}c@{}}{Geometrical estimation} & 
\multicolumn{2}{@{}c@{}}{Depletion\footnotemark[2] (\AA)}\\
& \multicolumn{2}{@{}c@{}}{Pore diameter\footnotemark[1] (\AA)} & & \\
\cmidrule{2-3}\cmidrule{4-5}%
 & MD & GO & MD & GO \\
\midrule
$a$-Pd              & -     & -        & -     & -  \\
$pa$-Pd$_{25\%}$    & 6.52	& 11.32	   & 3.16  & 7.78 \\
$pa$-Pd$_{50\%}$    & 10.53	& 12.41	   & 7.79  & 7.94 \\
$pa$-Pd$_{62.5\%}$  & 16.34	& 17.14	   & 8.86  & 9.60 \\
$pa$-Pd$_{75\%}$    & 16.67	& 18.83	   & 10.30 & 12.73 \\
\botrule
\end{tabular*}
\footnotetext[1]{Average diameter of all the samples for every porosity.}
\footnotetext[2]{Calculated from the depletion of the average $G(r)$ in Figure 2.}
\end{minipage}
\end{center}
\end{table}

\begin{figure}[ht]%
\centering
\includegraphics[width=\textwidth]{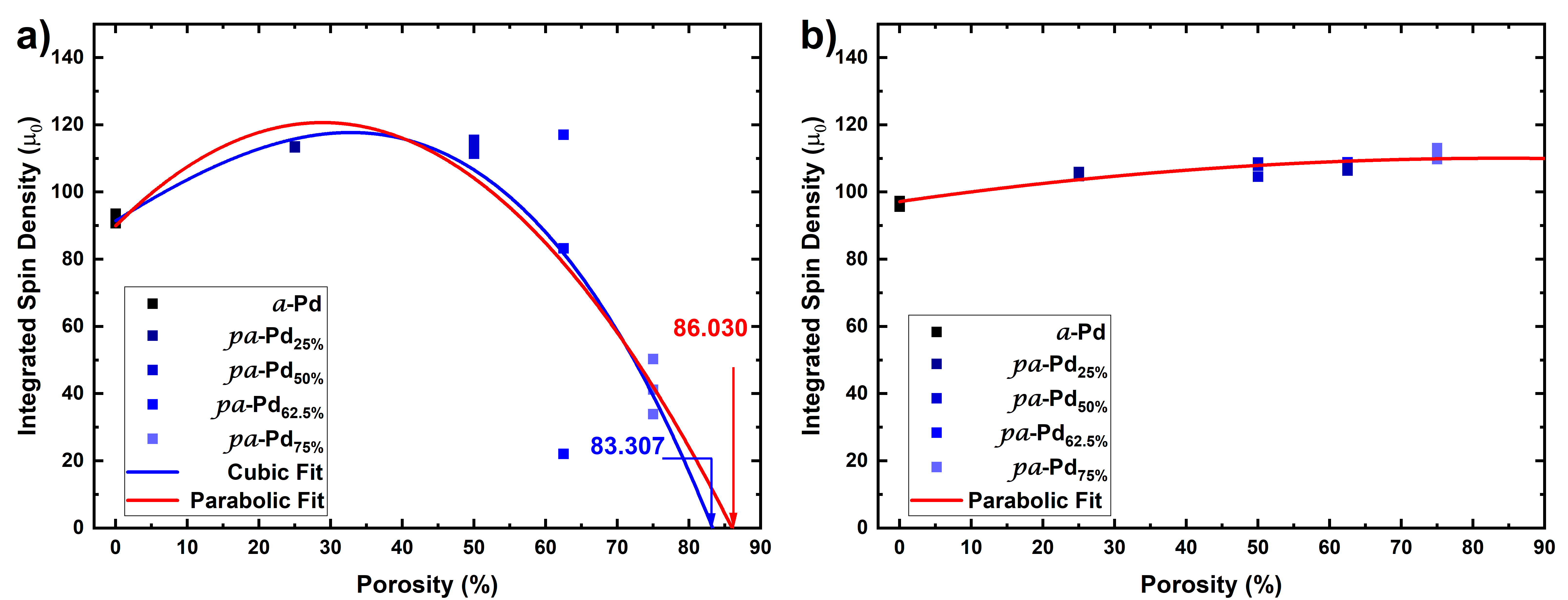}
\caption{Behavior of the integrated spin density as a function of the porosity. a) after the MD runs. b) after the GO runs. Color online.}\label{fig5}
\end{figure}

Once the MD and GO processes terminated, we investigated the electronic structure and the magnetism of the samples after each process. After the MD runs, the integrated spin density displays a systematic increase with porosity, with a maximum at about 40\% and then diminishes to practically zero for porosities larger than 80\%, as shown in Figure \ref{fig5} a). Figure \ref{fig5} b) shows the behavior of the integrated spin density as a function of porosity after the GO jobs terminated. For porosities greater than 50\% large voids were observed and a magnetic behavior different from the one obtained after the MD runs was observed.\\

These results merit a more careful analysis. Since the number of atoms in every cell is the same, 216, Figure \ref{fig5} a) indicates that the orientation of the magnetic moments becomes more aligned as the 40\% porosity region is approached, whereas before and after 40\% this orientation does not reinforce the magnetism and a decrease is observed when departing from this region. This indicates an increasing randomness in the magnetic moments before and after due to the MD processes that the supercells underwent, and as the middle of the porosity range is approached the tendency for the magnetic moments is to become more parallel. This phenomenon is observed with clarity when we look at the results after the GO processes. Figure \ref{fig5} b) shows that when the geometry optimization is conducted for each sample, the way to minimize the energy is by aligning all the magnetic moments in each atom, and since the number of atoms is constant the net magnetism is practically the same for each cell due to the fact that each atom contributes the same amount to the total magnetic moment. In our \textit{expanded-lattice} method, the total number of atoms in a sample does not change and that explains why the magnetism is practically constant after the GO for each cell no matter what the porosity value is. If we could generate a supercell with a 100\% porosity (infinite volume) then the number of atoms would still be 216 and the net magnetic moment would remain the same for the cluster that agglutinates all 216 atoms.\\

\begin{figure}[ht]%
\centering
\includegraphics[width=\textwidth]{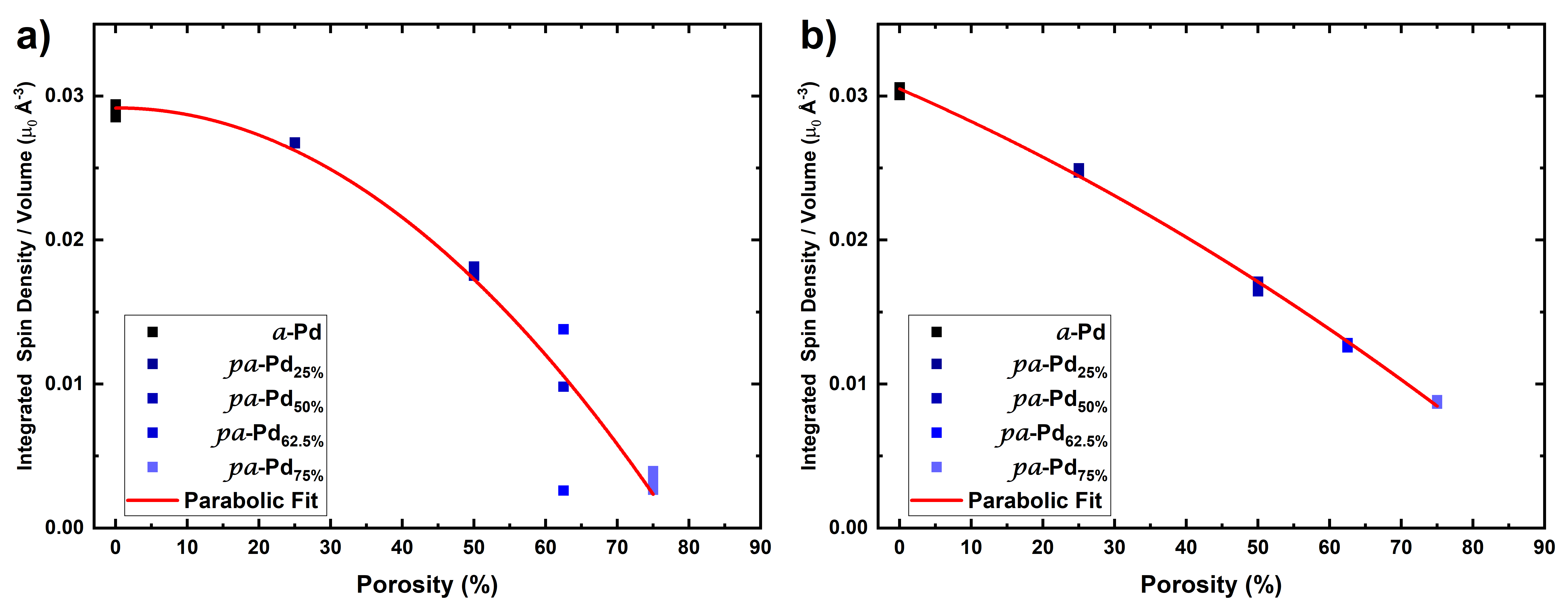}
\caption{Behavior of the integrated spin density per cell volume as a function of the porosity. a) after the MD runs. b) after the GO runs. Color online.}\label{fig6}
\end{figure}

If we consider the magnetism per unit volume this behavior can be clearly seen, as in Figures \ref{fig6} a) and b). As the volume tends to infinity with the atoms in the supercells remaining constant, then the magnetism per unit volume would tend to zero, as shown. The technological implications should be considered: If the total number of atoms in the solid diminishes due to the creation of porosity, the magnetism would also diminish and in the 30 to 40\% region the non-equilibrium MD process would generate the maximum of magnetism congruent with the number of atoms in the porous sample. After the GO process the total magnetic moment would be constant but if the number of atoms is scarce then the total magnetism would be also small and not very practical. The best thing to do is to create a geometry-optimized porous sample with a large number of atoms to take advantage of both aspects. A compromise is needed between the total number of atoms, porosity and equilibrated structures.

\section{Possible applications: A light-weight magnet.}\label{Possible}

The possibility of constructing magnets with medium to strong magnetism by developing amorphous porous palladium exists. The structures that are generated after the molecular dynamics processes indicate that the non-optimized samples behave in a parabolic manner and their magnetism is similar to that obtained after the geometry optimization procedure. This magnetism goes to zero for the highly porous cells. Evidently, the structures after the GO runs, both magnetic and topological, are different from the corresponding structures after the MD processes. This is due to the search for a local stable topology during the GO processes whereas no local equilibrium can be invoked after the MD jobs. The porosity affects seriously these properties as opposed to the non-porous samples.\\

If a process like chemical dissolution (dealloying) is invoked to generate porous palladium it would lead to a stressed atomic topology that would relax to a more stable structure in the course of time. Therefore, as these structures evolve, the magnetism would change and a mechanism to make them stable is required to avoid the changes in the magnetic value. The MD arrangements are the ones that would evolve more readily since they are not at (local) equilibrium, whereas the ones generated after the GO processes would have acquired a (local) stability. How long would the arrangements after MD remain static? It is hard to know with the computer simulation techniques applied in this work. We can only hope that the MD structures would remain static long enough to be useful.\\

Experimentalists are ingenious and depending on the magnetism desired they will be able to develop the necessary processes to design the adequate structure to produce the correct magnet, by dealing with the porosity and with the equilibrium characteristics. However, this may not be a trivial task and researchers would have to be inventive to accomplish the desired results. Since magnetic materials are used practically everywhere, the quest for the correct procedure may be worth searching for. For some applications the ratio (magnetism / weight) needs to be optimized since an excess of weight may hinder its use due to “fueling” considerations.  Therefore, despite the high cost of palladium in the market, it may be worthwhile to investigate the option presented here.

\section{Conclusions}\label{Conclusions}

Magnetism pervades our lives. Since primitive man discovered the properties of lodestone, the search for an understanding of the origin of magnetism in materials has been pursued. To date, several types of magnetism are reported and there seems to be more in the future \citep{musgraves_springer_2019, waseda_structure_1980}. Here we have computationally researched the properties of palladium with defects, in order to investigate their magnetic properties. We found that amorphicity generates magnetism in the material and the porosity in the amorphous also does, fostering the possible development of light-weight magnets of potential application in our environment. The nature of magnetism in $pa$-Pd is being investigated and the possible existence of magnetic domains in the realm of sub-nano dimensions is appealing, since this would lead to consider the formation of magnetic regions of very small sizes that contain only a few atoms, an interesting task.

\backmatter

\bmhead{Acknowledgments}

I. Rodríguez thanks PAPIIT, DGAPA-UNAM for his postdoctoral fellowship. D. Hinojosa-Romero acknowledges Consejo Nacional de Ciencia y Tecnología (CONACyT) for supporting his graduate studies. A. A. Valladares, R. Valladares and A. Valladares thank DGAPA-UNAM (PAPIIT) for continued financial support to carry out research projects under Grants No. IN104617 and IN116520. M.T. Vázquez and O. Jiménez provided the information requested. A. Lopez and A. Pompa assisted with the technical support and maintenance of the computing unit at IIM-UNAM. Simulations were partially carried at the Computing Center of DGTIC-UNAM under project LANCAD-UNAM-DGTIC-131.

\section*{Declarations}

\begin{itemize}
\item Funding: Financial support was received from DGAPA-UNAM (PAPIIT) under projects No. IN104617 and IN116520.
\item Conflict of interest/Competing interests: The authors have no competing interests to declare that are relevant to the content of this article.
\item Ethics approval: Not applicable.
\item Consent to participate: Not applicable.
\item Consent for publication: The authors gave their consent for the publication of this manuscript.
\item Availability of data and materials: The datasets analysed during the current study are available from the corresponding author on reasonable request. 
\item Code availability: Code Correlation can be found in the GitHub repository: https://github.com/Isurwars/Correlation (DOI: 10.5281/zenodo.5514113).
\item Authors' contributions: Ariel A. Valladares, Alexander Valladares and Renela M. Valladares conceived this research and designed it with the participation of Isa\'{i}as Rodr\'{i}guez and David Hinojosa-Romero. All the simulations were done by Isa\'{i}as Rodr\'{i}guez. and David Hinojosa-Romero. All authors discussed and analyzed the results. Ariel A. Valladares wrote the first draft and the other authors enriched the manuscript.
\end{itemize}

\bibliography{sn-bibliography}% common bib file

\end{document}